# A Fair and Efficient Packet Scheduling Scheme for IEEE 802.16 Broadband Wireless Access Systems

Prasun Chowdhury[1], Iti Saha Misra[2]

Electronics and Telecommunication Engineering
Jadavpur University, Kolkata, India
[1]prasun.jucal@gmail.com, [2]iti@etce.jdvu.ac.in

### *Abstract*

This paper proposes a fair and efficient QoS scheduling scheme for IEEE 802.16 BWA systems that satisfies both throughput and delay guarantee to various real and non-real time applications. The proposed QoS scheduling scheme is compared with an existing QoS scheduling scheme proposed in literature in recent past. Simulation results show that the proposed scheduling scheme can provide a tight QoS guarantee in terms of delay, delay violation rate and throughput for all types of traffic as defined in the WiMAX standard, thereby maintaining the fairness and helps to eliminate starvation of lower priority class services. Bandwidth utilization of the system and fairness index of the resources are also encountered to validate the QoS provided by our proposed scheduling scheme.

## 1. Introduction

Wireless networks are generally less efficient and unpredictable compared to wired networks, which make Quality of Service (QoS) provisioning a bigger challenge for wireless communications. The wireless medium has limited bandwidth, higher packet error rate, and higher packet overheads that altogether limit the capacity of the network to offer guaranteed QoS. In response to increasing QoS challenge, the IEEE 802.16 standard, also known as Worldwide Interoperability for Microwave Access (WiMAX), has emerged as the strongest contender for Broadband wireless technology with the promises to offer guaranteed QoS to wireless users. WiMAX is a technology aimed at providing last-mile wireless broadband access at a cheaper cost. The "last mile" is the final leg of delivering connectivity from the service provider to the customer [1]. This leg is typically seen as an expensive undertaking because of the considerable costs of wires and cables. The core of WiMAX technology is specified by the IEEE 802.16 standard that provides specifications for the Medium Access Control (MAC) and Physical (PHY) layers. The term WiMAX was created by the WiMAX forum that promotes conformance and interoperability of the standard.

In WiMAX network, traffic from the Base Station (BS) to the Subscriber Station (SS) is classified as downlink traffic while that from the SS to the BS is classified as uplink traffic. A scheduling algorithm implemented at the BS has to deal with both uplink and downlink traffic. In some cases, separate scheduling algorithms are implemented for the uplink and downlink traffic. Typically, a Call Admission Control (CAC) procedure is also implemented at the BS that ensures the load supplied by the SSs can be handled by the network [1-5]. A CAC algorithm ensures admission of a SS into the network if it can satisfy minimum Quality of Service (QoS) requirements and at the same time QoS of existing SSs will not deteriorate. The performance of the scheduling algorithm for the uplink traffic strongly depends on the CAC algorithm.

Scheduling is a critical component of Worldwide Interoperability of Microwave Access (WiMAX) impacting significantly on its performance. Scheduling schemes helps in providing service guarantees to heterogeneous classes of traffic with different QoS requirements. In addition to scheduling, bandwidth request and bandwidth allocation mechanisms also play crucial roles in QoS provisioning for WiMAX. In general, a scheduler for WiMAX needs to be simple, efficient, fair, scalable, and have low computational complexity. It should also be able to protect against misbehaving flows and provide decoupling and necessary bounds on throughput and delay performance.

Packet scheduling [5-14] is the process of resolving contention for shared resources in a network. The process involves allocating bandwidth among the users and determining their transmission order. Scheduling algorithms for a particular network need to be selected based on the type of users in the network and their QoS requirements. QoS requirements vary depending on the type of application/user. For real-time applications such as video conferencing, voice chat and audio/video streaming, delay and delay jitter are the most important QoS requirements. Delay jitter is the inter-packet arrival time at the receiver and is required to be reasonably stable by the real-time applications. On the other hand, for non-real time applications such as file transfer protocol (FTP), throughput is the most important QoS requirement. Some applications, such as web-browsing and email do not have any QoS requirements. In a network, different types of applications, with diverse QoS requirements can co-exist. A task of a scheduling algorithm in a multi-class network is to categorize the users into one of the pre-defined classes. Each user is assigned a priority taking into account its QoS requirements. Subsequently, bandwidth is allocated according to the priority of the users as well as ensuring that fairness between the users is maintained. Fairness refers to the equal allocation of network resources among the various users operating in both good and bad channel states. In this paper, fairness is quantified using Jain's Fairness Index [15]. In addition to it, Bandwidth Utilization of the system is also encountered to estimate whether precious bandwidth will get wasted by SS lying in a bad channel state.

Packet scheduling algorithms are implemented at both the BS and SSs. A scheduling algorithm at the SS is required to distribute the bandwidth allocation from the BS among its connections. A scheduling algorithm at the SS is not needed if the BS grants bandwidth to each connection of the SS separately i.e. the Grant per Connection (GPC) procedure is followed. If the Grant per Subscriber Station (GPSS) procedure is followed, the scheduling algorithm at the SS needs to decide on the allocation of bandwidth among its connections. The scheduling algorithm implemented at the SS can be different than that at the BS [8].

The focus of our work is on scheduling algorithms for the uplink traffic in WiMAX i.e. traffic from the SSs to the BS. Uplink packet scheduling is a more challenging task than downlink packet scheduling as all the necessary information of SSs such as queue size for the uplink scheduling are not available. An uplink algorithm at the BS has to coordinate its decision with all the SSs where as a downlink algorithm is only concerned in communicating the decision locally to the BS.

K. Wongthavarawat *et al.* propose a hybrid scheduling algorithm in [1] that combines Earliest Deadline First (EDF), Weighted Fair queuing (WFQ) and First in First out (FIFO) scheduling algorithms. The overall allocation of bandwidth is done in a strict priority manner i.e. all the higher priority SSs are allocated bandwidth until they do not have any packets to send. The EDF scheduling algorithm is used for SSs of the rtPS class, WFQ is used for SSs of the Non-Real Time Polling Service (nrtPS) class and FIFO for SSs of the Best Effort (BE) class. Besides the scheduling algorithm, an admission control procedure and a traffic policing mechanism are also proposed. All these components together constitute the proposed QoS architecture. A drawback of this algorithm is that lower priority SSs will essentially starve in the presence of a large number of higher priority SSs due to the strict priority overall bandwidth allocation.

J. Lin *et al.* [8] propose architecture called Multi-class Uplink Fair Scheduling Structure (MUFSS) to satisfy throughput and delay requirements of the multi-class traffic in WiMAX. The proposed scheduling discipline at the BS is Modified Weighted Round Robin (MWRR), although details of the modifications to the Weighted Round Robin (WRR) discipline are not provided by the authors. The model is based on Grant per Subscriber Station (GPSS) bandwidth grant mode and thus schedulers are implemented at the SSs to distribute the bandwidth granted among their connections. At the SS, Modified WFQ (MWFQ) is used for Unsolicited Grant Service (UGS) and Real-Time Polling Service (rtPS) connections, MWRR is used for nrtPS connections and FIFO is used for BE connections.

K. Vinay *et al.* [9] propose a hybrid scheme that uses EDF for SSs of the rtPS class and WFQ for SSs of the nrtPS and BE classes. This algorithm differs from the one in [1] in a couple of ways. First, the WFQ algorithm is used for SSs of both nrtPS and BE classes. Secondly, the overall bandwidth allocation is not done in a strict priority manner. Although the details of overall bandwidth allocation are not specified, it is briefly mentioned that the bandwidth is allocated among the classes in a fair manner. Since SSs of the BE class do not have any QoS requirements, using a computationally complex algorithm such as WFQ for them is not needed. Here author made the comparative study of the scheduling algorithms implemented in GPSS and GPC and found that GPSS gives better end-to-end delay.

M.Settembre *et al.* [10] propose a hybrid scheduling algorithm that uses WRR and Round Robin (RR) algorithms with a strict priority mechanism for overall bandwidth allocation. In the initial portion of the algorithm, bandwidth is allocated on a strict priority basis to SSs of the rtPS and nrtPS classes only. After that the WRR algorithm is used to allocate bandwidth among SSs of rtPS and nrtPS classes until they are satisfied. If any bandwidth remains, it is distributed among the SSs of the BE class using the RR algorithm. This algorithm starves lower priority SSs in the presence of a large number of higher priority SSs. The algorithm can also result in low fairness among SSs as it selects SSs with the most robust burst profiles first.

J. SUN *et al.* [11] proposed that the scheduler inside the BS may have only limited or even outdated information about the current state of each uplink connection due to the large Round Trip Delay (RTD) and possible collision occurred in the uplink channel transmission. So there is a need of an additional scheduler in each SS to reassign the received transmission opportunities among different connections. Since the uplink traffic is generated at SS, the distributed scheduler is able to arrange the transmission based on the up-to-date information and then provide QoS guarantee for its connections. But here the proposed algorithm is suffered by a problem called as *starvation of lower priority class services*.

It is observed that the existing wire line and wireless schedulers do not perform very well with respect to different traffic classes defined in WiMAX. In addition, each of this traffic classes has a different scheduling requirement and, consequently, it has become necessary to design appropriate hybrid scheduling frameworks. So in the proposed method the uplink traffic is scheduled based on current queue information at SS similarly in the way proposed in [11]. But a different hybrid algorithm has been implemented at the SS scheduler which helps to eliminate the starvation of lower priority class services also maintains proper fairness and bandwidth utilization of the system even at lower traffic intensity. In the proposed method the BS scheduler can guarantee the minimum bandwidth for each service flow and ensure fairness and QoS in distributing excess bandwidth among all connections. At the same time, the scheduler in SS can provide differentiated and flexible QoS support for all of the four scheduling service types. In this paper EDF algorithm is applied for rtPS class of services and Deficit Fair Priority Queue (DFPQ) algorithm found in literature [12],[13] is applied for nrtPS and BE class of services. It can both reduce the delay of real-time applications and guarantee the throughput of non-real-time applications such as nrtPS and BE.

The rest of the paper is organized as follows; The QoS related features of IEEE 802.16 standard are discussed in section 2. Then the proposed scheduling algorithms are introduced in section 3. Section 4 provides simulation and performance analysis. Finally, this paper is ended up with the conclusions drawn in section 5.

## 2. QoS Features of IEEE 802.16

The first version, known as 802.16, was completed in October 2001. It specified a Single Carrier (SC) air interface for fixed point-to-multipoint (PMP) BWA systems operating between 10-66 GHz. The second amendment, 802.16a, was published in January 2003. It extends the physical environment towards lower frequency bands below 11 GHz. The next approved version is 802.16d, which is published in June 2004 and also known as FIXED WiMAX (802.16-2004). It incorporates all the previous versions to provide fixed BWA. In 2005, IEEE undertakes the standardization of 802.16e, which is expected to support full mobility up to 70-80miles/sec [6]. Four service types are defined in IEEE 802.16d-2004 (Fixed) standard, which includes UGS (Unsolicited Grant Service), rtPS (Real-time Polling Service), nrtPS (Non Real-time Polling Service), and BE (Best Effort). The UGS is designed to support real-time service flow that generates fixed-size data periodically, such as T1/E1, VoIP without silence suppression [1], [6]. The rtPS is designed to support real-time service flow that generates variable size data, such as video streaming services while the nrtPS deals with FTP in similar manner. The BE perform tasks related to e-mail and web browsing. The guaranteed delay aspect is also taken care in video streaming and VoIP.

Since IEEE 802.16 MAC protocol is connection oriented, the application first establishes the connection with the BS as well as the associated service flow (UGS, rtPS, nrtPS or BE). BS will assign the connection with a unique connection ID (CID) [1], [6]. All packets from the application layer in the SS are classified by the connection classifier based on CID and are forwarded to the appropriate queue. So, the scheduler inside the BS has outdated information about the current state of each uplink connection due to the large Round Trip Delay (RTD) and possible collision occurred in the uplink channel transmission [14].

## 3. Proposed Scheduling Algorithm

In the proposed method, the uplink bandwidth allocation at BS is done based on the per connection requests from SSs. Because a SS may have multiple connections at the same time, the bandwidth request messages should report the bandwidth requirement of each connection in SS. After that the allocated bandwidth per connection is pooled together and granted to each SS. Then SS re-distribute the received transmission opportunities among its connections according to their QoS requirement. Therefore an additional scheduler is needed in each SS to reassign the received transmission opportunities among different connections. Since the uplink traffic is generated at SS, the distributed scheduler is able to arrange the transmission based on the up-to-date information and then provide tight QoS guarantee for its connections.

Since the BS scheduler has limited information on the traffic generated at SS, the computing of bandwidth allocation should just consider the bandwidth request and reservation for each connection.

Let $BWMIN_i$ denote the minimum reserved bandwidth for connection i, and $BWREQ_i$ represent the bandwidth currently demanded by the connection i. Since the connection will never get more resources than it has requested, the bandwidth actually allocated ($BWALLOCATE_i$) during this phase is

$$BWALLOCATE_i = \min \{BWMIN_i, BWREQ_i\} \quad (1)$$

For rtPS and nrtPS, $BWMIN_i$ is specified by the QoS parameter termed Minimum Reserved Traffic Rate. Clearly, to guarantee the contracted bandwidth, the sum of minimum reserved bandwidth for all the connections should not exceed the available bandwidth B. After each connection gets its guaranteed bandwidth, if there is still excess uplink bandwidth remained, BS scheduler should distribute the residual bandwidth in proportion to the pre-assigned connection weight. The algorithm in this phase can be described as:

```
BWREMAIN = B - ∑ BWALLOCATEi;
               i
While BWREMAIN > 0
{
     If (BWALLOCATEi < BWREQi)
     {
                                  n
          BWADDi = BWREMAIN*Wi / ∑  Wk;
                                k=1
          BWALLOCATEi = BWALLOCATEi + BWADDi;
          BWREMAIN = BWREMAIN - BWADDi;
      }
}
```

Where BWREMAIN is the remaining bandwidth, $BWADD_i$ is the amount of excess bandwidth allocated to connection queue i and $W_i$ is the weight of connection queue i. Now the allocated bandwidth per connection is pooled together and granted to each SS.

SS scheduler will select the packet to be transmitted from the highest priority queue. The priority of the queue is maintained in the following way UGS > rtPS > nrtPS > BE.

### 3.1 Scheduling algorithm for UGS queues

UGS generate fixed size data packets on a periodic basis. This service has a critical delay and delay jitter requirement. So, SS scheduler firstly guarantees the bandwidth for UGS queues.

### 3.2 Scheduling algorithm for rtPS queues

For rtPS service, each packet entering the rtPS queues should be marked with a delivery deadline equal to t + tolerated delay, where t is the arrival time and tolerated delay is the Maximum Latency for such a service flow. The packet with smaller deadline will be transmitted earlier. This greatly reduces the end-to-end delay of rtPS service.

### 3.3 Scheduling algorithm for nrtPS and BE queues

For nrtPS and BE services Deficit Fair Priority Queue (DFPQ) algorithm found in literature [12], [13] is employed. DFPQ is almost similar to Deficit Round Robin (DRR) algorithm. This algorithm has been applied because of the following reasons.

i. The algorithm is mostly suited for datagram networks where packet sizes vary.

ii. Since this algorithm requires accurate knowledge of packet size, it is suitable for the uplink traffic at SS scheduler.

*iii.* The algorithm is flexible enough as it allows provision of quanta of different sizes depending on the QoS requirements of the SSs. With this algorithm employed, SS scheduler can guarantee the minimum bandwidth for every non real time services such as nrtPS and BE connection and hence maintain an acceptable throughput. *Thereby eliminate starvation of lower priority service classes.*

In each service round, the nrtPS queue is served first until its assigned bandwidth finds deficit, and then BE service flow queue gets a chance to be served. Similar to [12], in DFPQ algorithm, a Quantum Q is assigned to each queue *i*. The quantum of a queue *i (Q[i])* represent the maximum number of bits that can be serviced in the first round. The scheduler visits each nonempty queue and determines the number of bandwidth requests in this queue. If there are more packets in the queue *i* after servicing *Q[i]* bits, the remaining amount of bits is stored in a queue state variable called *Deficit Counter (DC[i])* and the scheduler moves on to serve the next non-empty queue. In subsequent rounds, the amount of bandwidth usable by this flow is the sum of *DC[i]* in the previous round added to *Q[i]*. The *Q*[*i*] is the Maximum Sustained traffic rate (rmax) of a certain service flow. In case *rmax* = 0 (BE service flow), *rmin* is used instead. As a result of using the quantum variable, connections with larger quantum get more service.

DFPQ algorithm is shown below. Here *Ltotal* is defined to be the remaining total capacity of the frame after servicing UGS and rtPS queues and *La* to be its remaining capacity. p(i,k) denotes $k^{th}$ packet of $i^{th}$ connection, i belongs to nrtPS and BE connections. The algorithm can be described as follows,

Each service queue i of nrtPS and BE is initialized with

```
La = Ltotal;
DC[i] = Q[i];

While La > 0
{
         While DC[i]>0 and p (i, k) is available
         {
             DC[i] = DC[i] - p(i,k);
             La =La – p(i,k);
             k++;
         }
         DC[i] = DC[i] + Q[i];
         i = (i % n) + 1;
}
Go to UGS scheduling for next frame
```

## 4. Simulation and Performance Analysis

To evaluate the effectiveness and efficiency of the proposed scheduler, the IEEE 802.16 MAC layer protocol is analyzed using MATLAB under version 7.3. A number of simulations are conducted in this section. At first, the simulation environment and parameters are described and simulation results are given along with the discussions.

## 4.1. Simulation Environment and Parameters

A TDD-OFDM system is used in our simulation with the MAC layer application parameters as shown in Table 1 and the network is configured as consists of one BS and multiple SSs as shown in Fig 1.

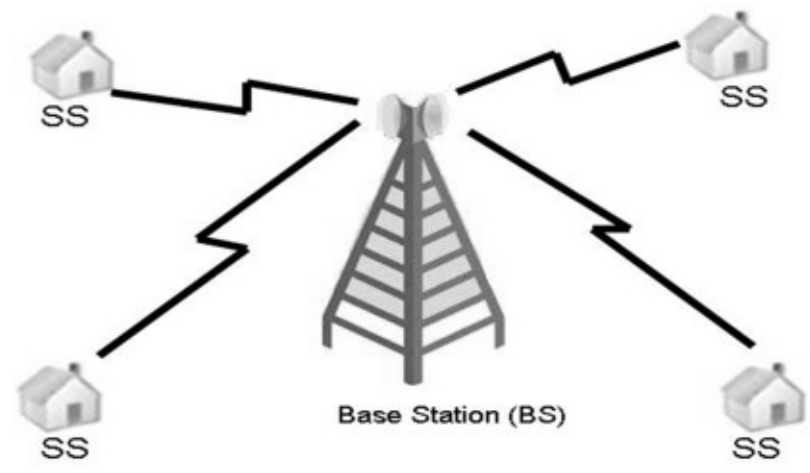

Figure1. Proposed model architecture

The frame duration is taken as 10ms with TDD duplex mode and the bandwidth is 4.3 MHz. The IEEE 802.16 standards have not specified values for the QoS parameters and we have assumed these values for the performance analysis.

Table1. MAC layer configuration parameters

| Service | Max Sustained Rate (kbps) | Min Reserved Rate (kbps) | Delay (ms) |
|---|---|---|---|
| UGS | 256 | - | - |
| rtPS | 1024 | 512 | 20 |
| nrtPS | 1024 | 512 | - |
| BE | - | 256 | - |

QoS parameters such as delay, delay violation rate, throughput, bandwidth utilization and fairness index are considered to validate our proposed scheduling scheme. Here, the delay violation rate is defined as the amount of packets whose delay is larger than the Maximum Latency to the total amount of packets that have been received from network interface. Also the bandwidth utilization is defined as the average ratio of used bandwidth to the total bandwidth and Fairness is quantified using Jain's Fairness Index (JFI) [15] as shown below.

$$JFI = \frac{(\sum_{i=1}^{n} r_i)^2}{n * \sum_{i=1}^{n} r_i^2} \qquad (2)$$

Where, $r_i$ is the data rate of application i.

## 4.2. Simulation Result and Discussions

In the simulation we consider 802.16 network consisting of one BS and four SSs with different traffic patterns. The first SS is configured with all types of traffic flows nominated as UGS_1, rtPS_1, nrtPS_1 and BE_1, the second SS has UGS_2, rtPS_2, nrtPS_2 and BE_2, the third SS has UGS_3, rtPS_3, nrtPS_3 and BE_3, the fourth SS runs UGS_4, rtPS_4, nrtPS_4 and BE_4. Two scenarios - with and without SS scheduler are simulated to study the effect of SS scheduler.

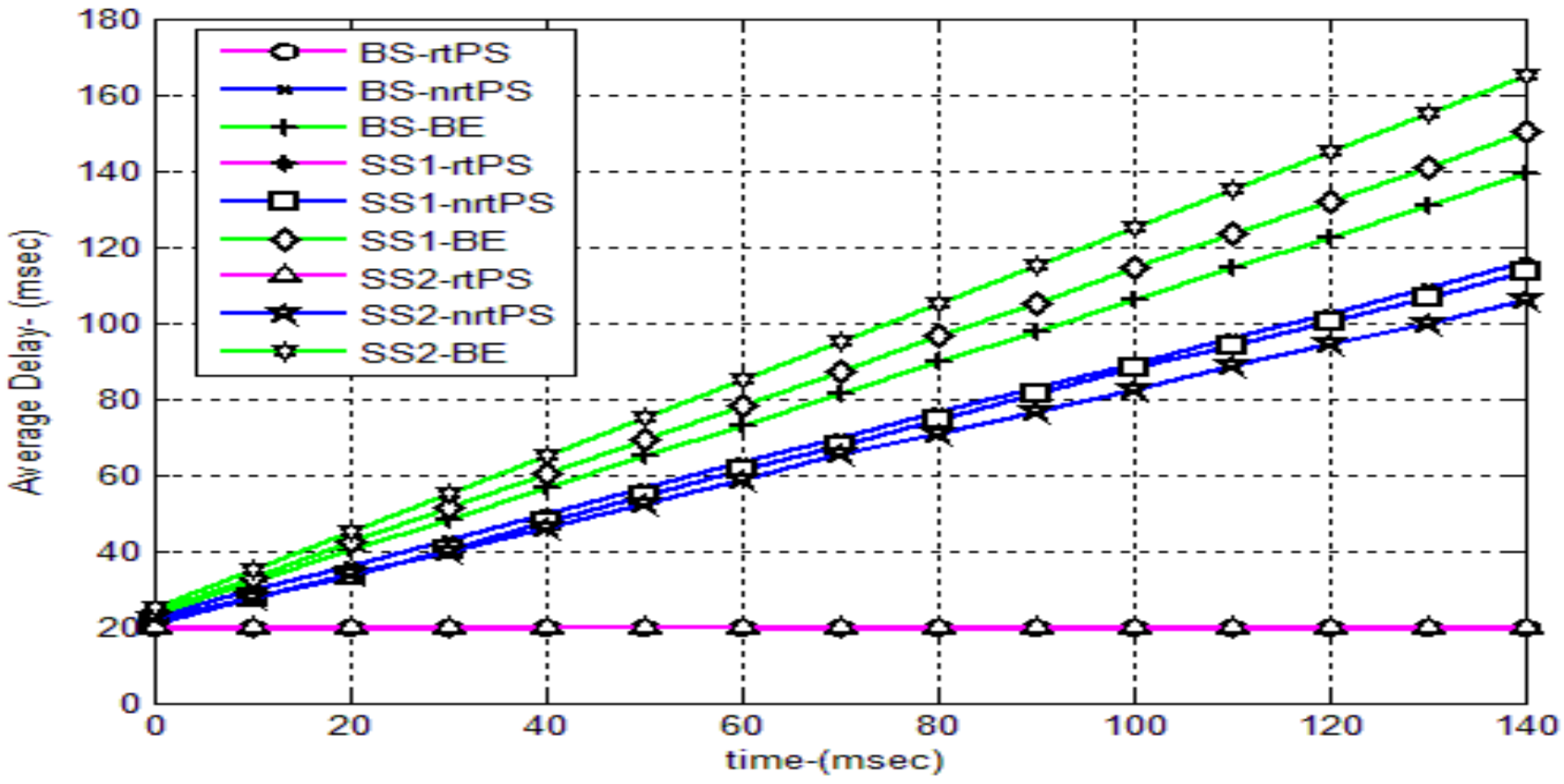

Figure 2.Service delay comparison

Our proposed method is compared with the method given in [11]. Here, "without SS scheduler" and "BS-(service class)" means that BS scheduler designates bandwidth to individual connection. On the other hand, "with SS-scheduler1" and "SS1-(service class)" means that SS scheduler designates bandwidth to individual connection in our proposed method. Again "with SS-scheduler2" and "SS2-(service class)" means that SS scheduler designates bandwidth to individual connection proposed in [11]. Here, service class refers to UGS, rtPS, nrtPS and BE**.** As UGS generates fixed size data packets on a periodic basis so delay is negligible and throughput is constant and hence it is not shown in our simulation result. Fig. 2 displays the end-to-end delay of different services with and without SS scheduler. The curves show that after SS scheduling, low priority service suffered longer delay. From rtPS, nrtPS to BE, the end-to-end delay increased with the service priority decreased. The fundamental requirement of QoS scheduling for IEEE 802.16 systems is achieved.

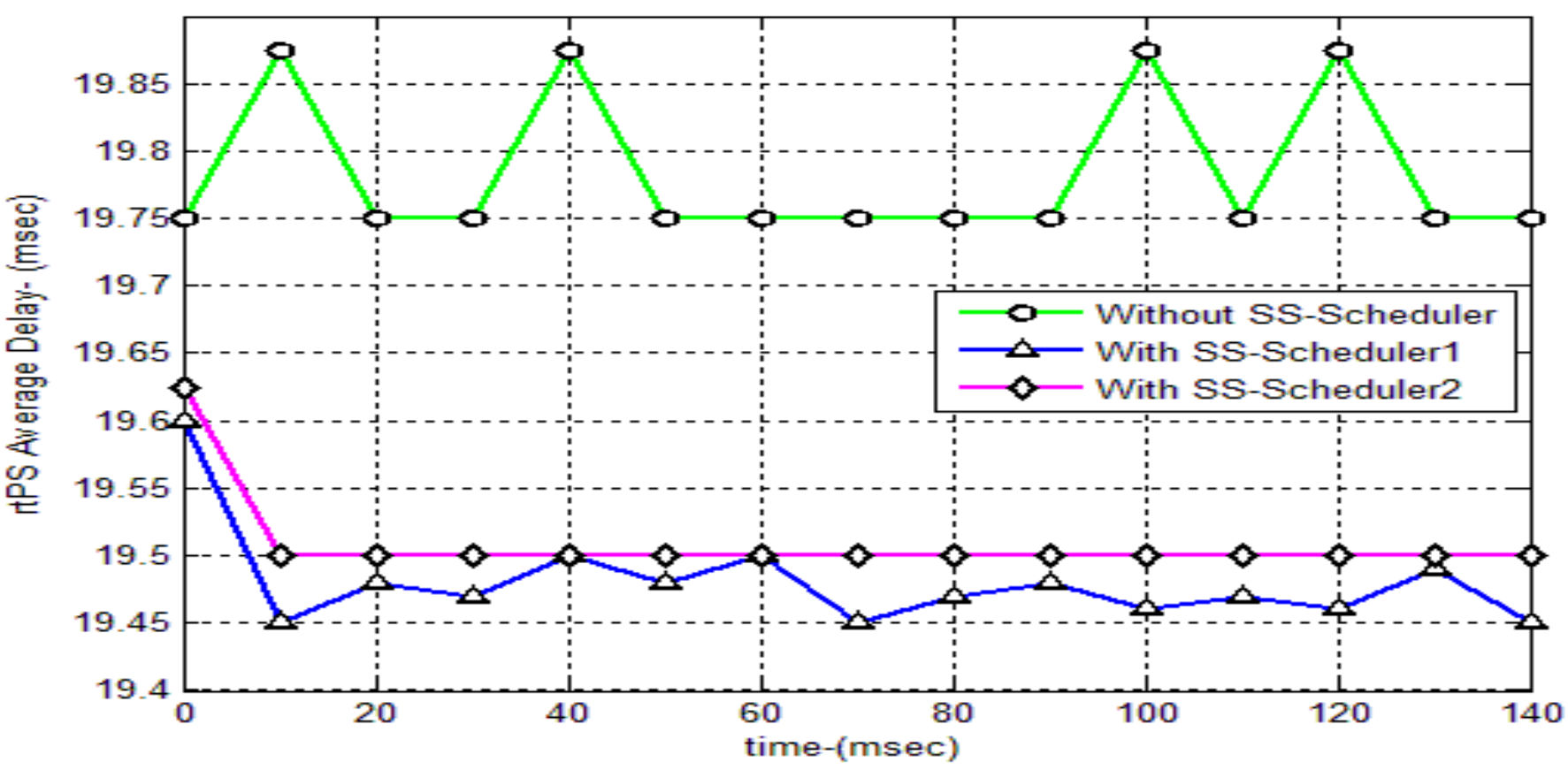

Figure 3.Service delay comparison of rtPS

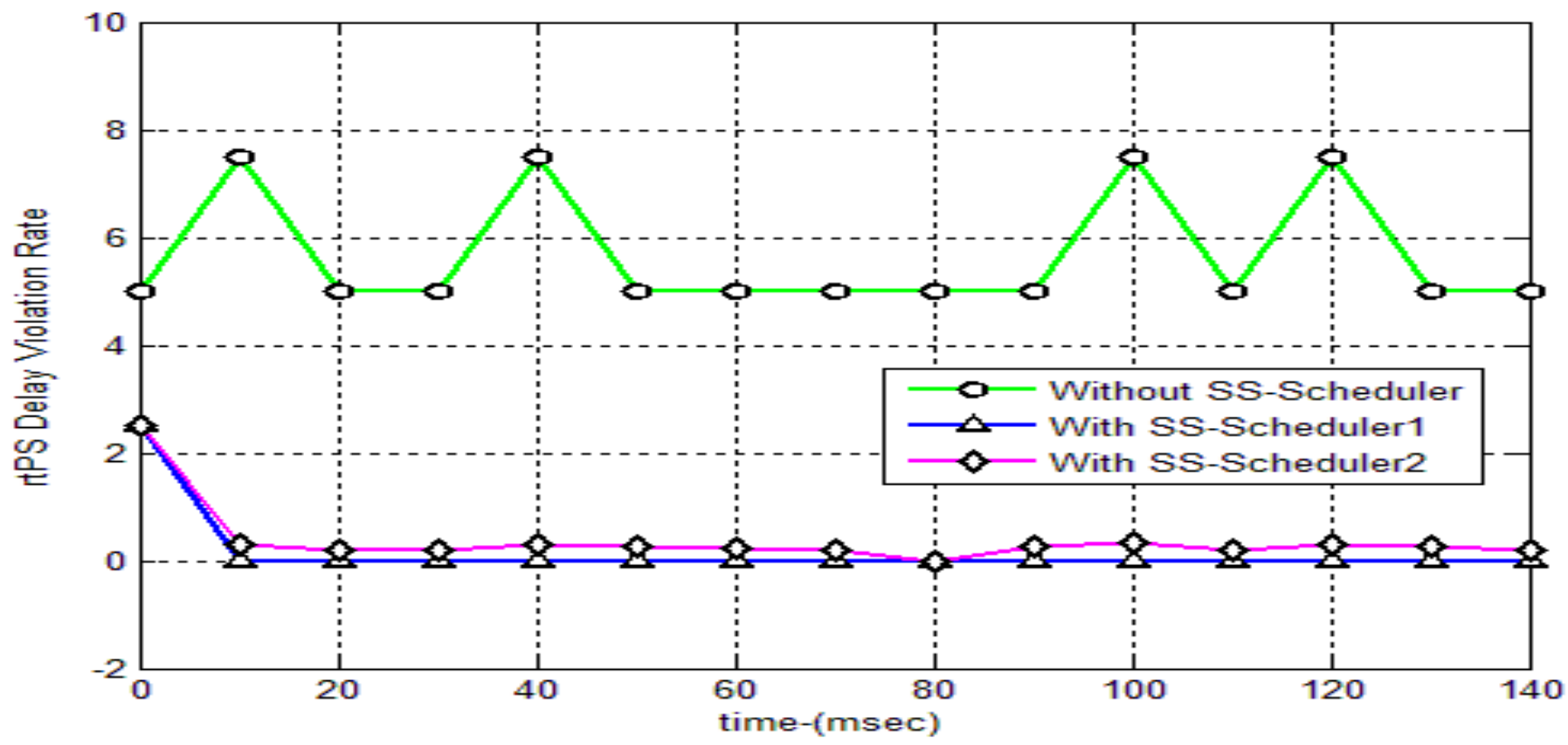

Figure 4.Percentage of packet drop comparison

To further demonstrate this benefit, we simulate the rtPS performance under the same number of background SS as given in Fig. 3. From Fig.4 we can see that the SS scheduler can effectively reduce the QoS violation rate of rtPS service flow.

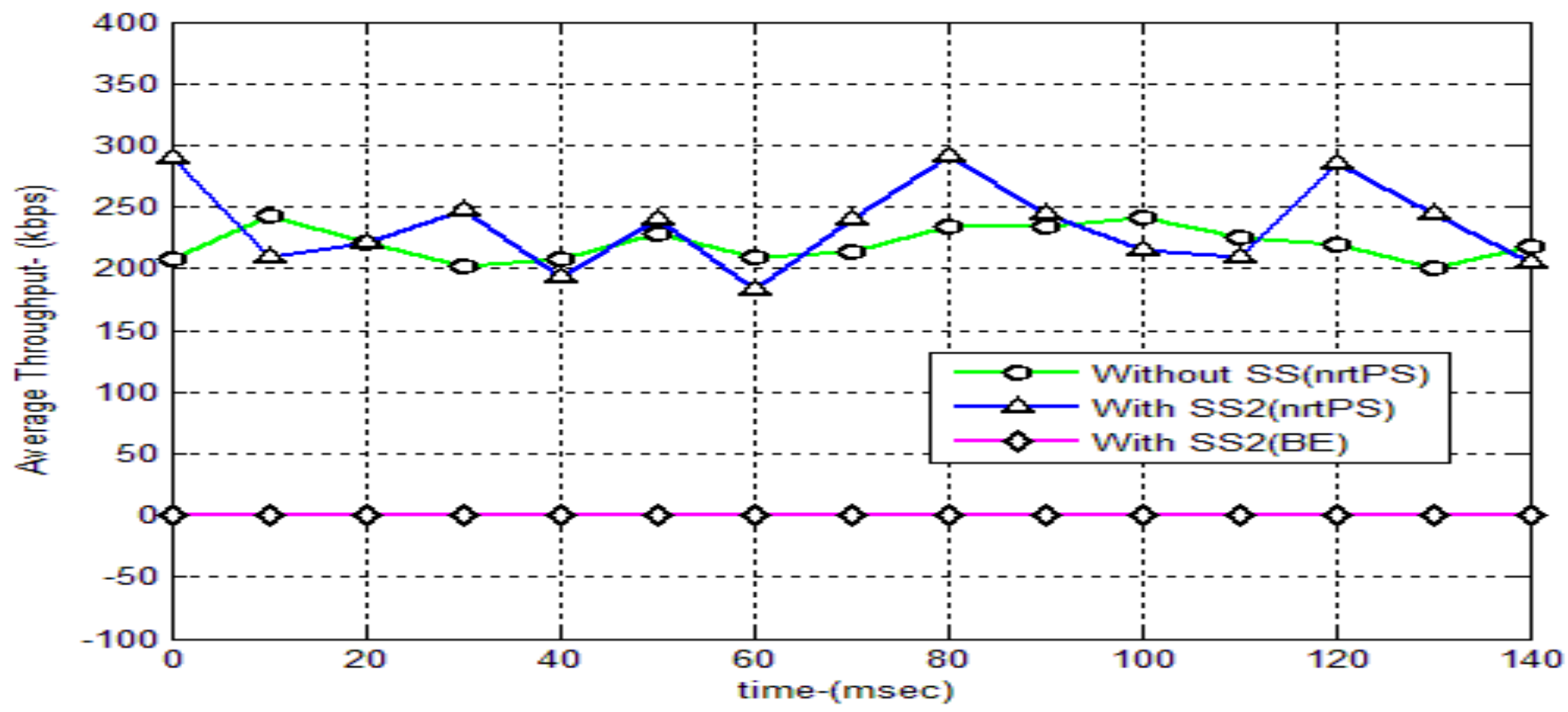

Figure 5.Throughput comparison proposed in [11]

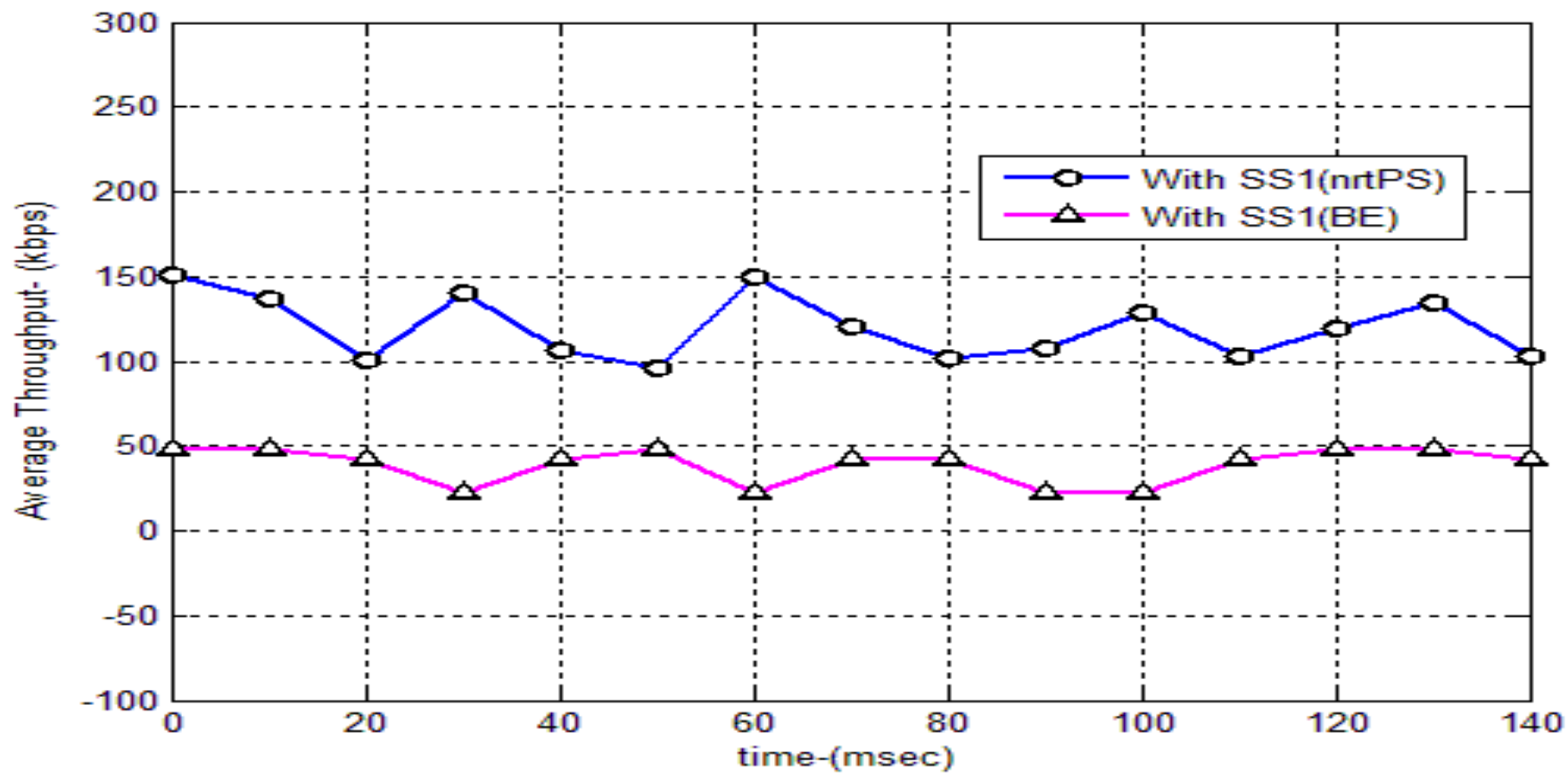

Figure 6.Throughput comparison in our proposed Strategy

By introducing DFPQ algorithm for nrtPS and BE services on priority order, SS scheduler can guarantee the throughput of nrtPS service as well as BE service. Though BE services do not require any QoS but in our proposed algorithm fairness is maintained for all types of service classes. But this fairness is not maintained in [11] where we got throughput of BE service is zero as shown in Fig.5. But our simulation result shows that SS scheduler guarantees the throughput of nrtPS service as well as BE service as shown in Fig.6 where throughput of BE service is greater than zero throughout simulation time, hence fairness of all the services is maintained and the problem of starvation of lower priority class services is eliminated.

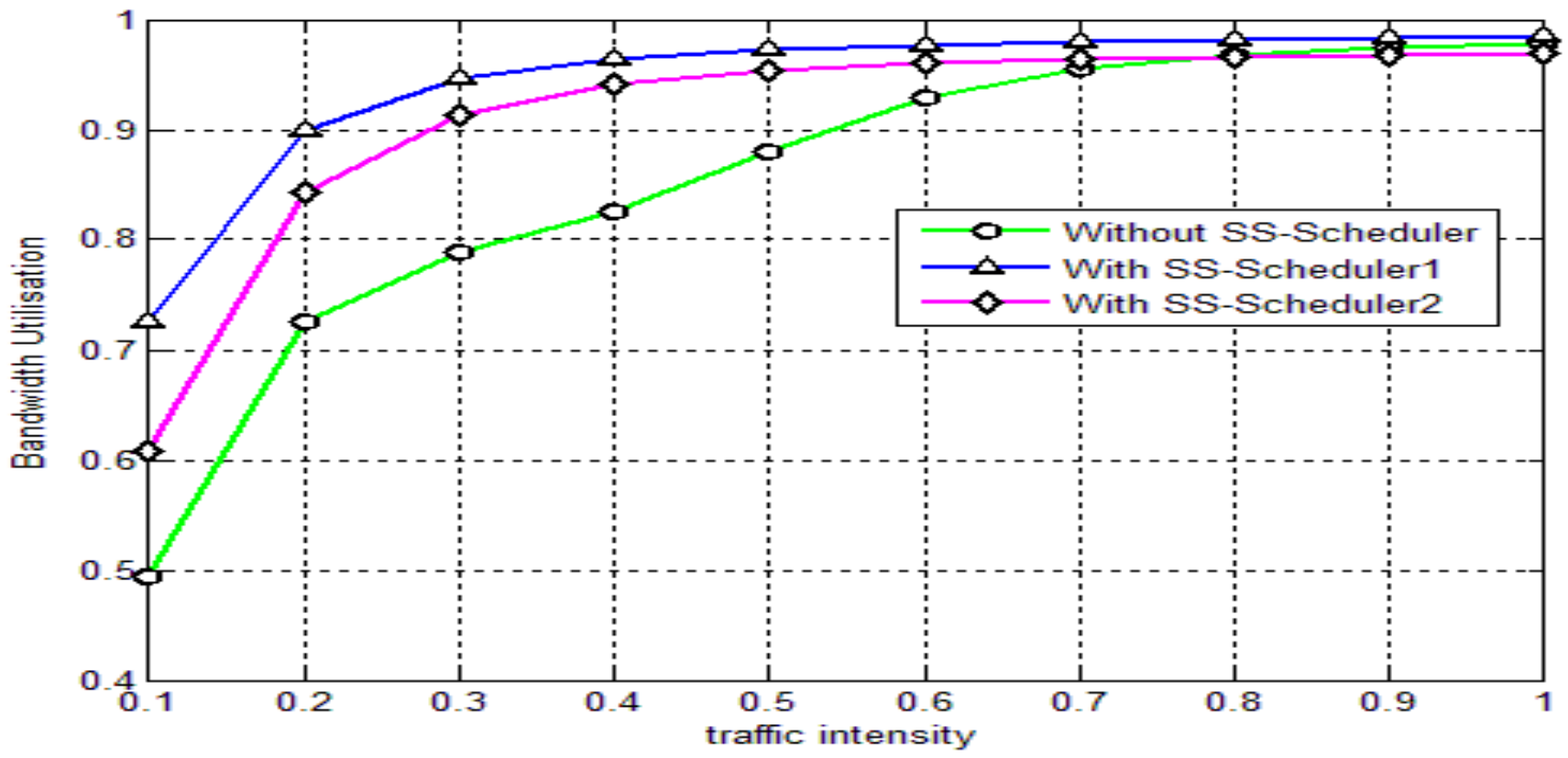

Figure 7.Bandwidth Utilization of the system

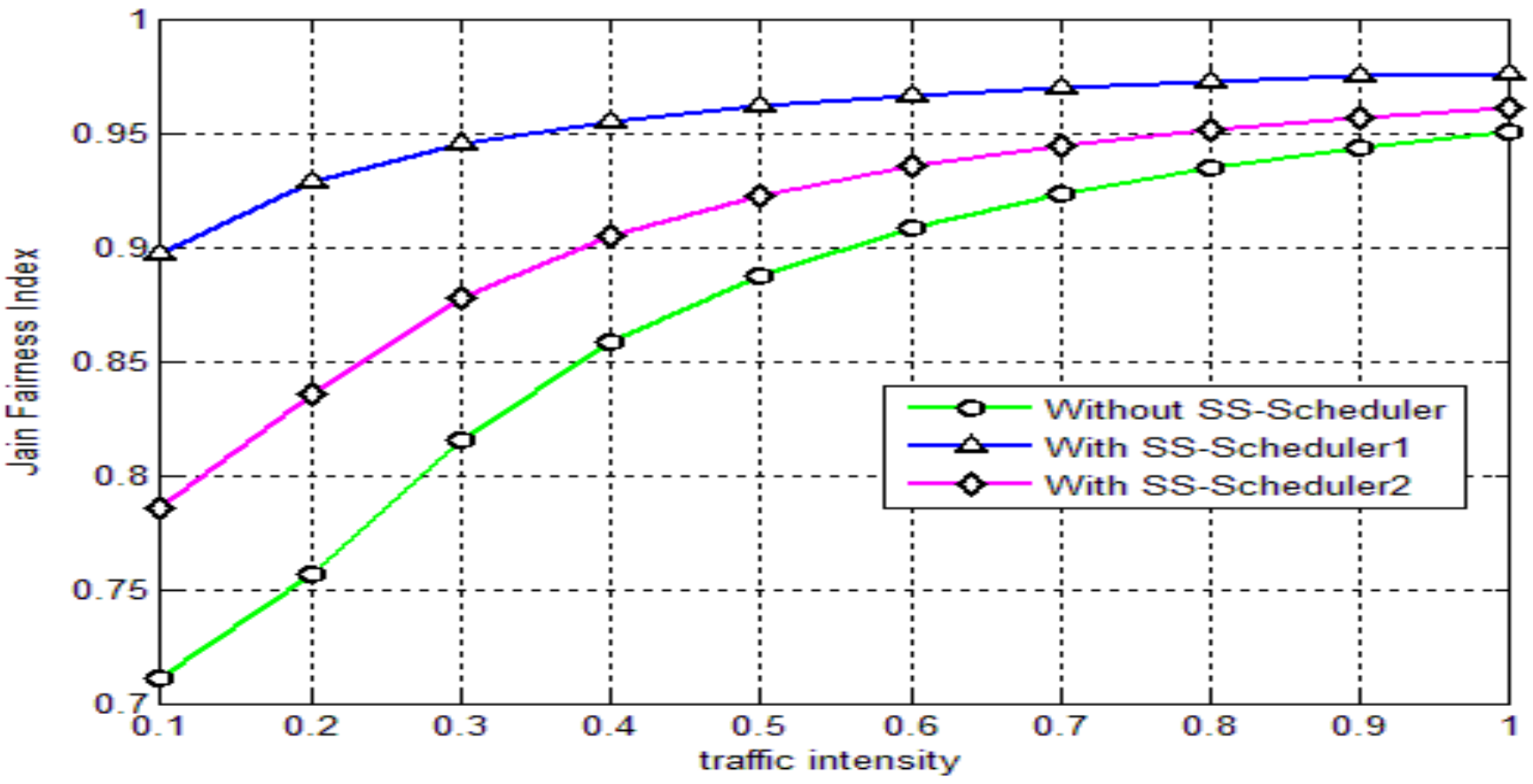

Figure 8.Jain's Fairness Index

Fig. 7 and 8 show bandwidth utilization of the system and fairness index for BS-Scheduler, SS1-Scheduler and SS2-Scheduler with different traffic intensity. Our results show that SS1-Scheduler performs well even at lower traffic intensity. As bandwidth is considered to be a limited resource in the network, so SS1-Scheduler will automatically improve the revenues of the service providers.

## 5. Conclusion

It has been confirmed in many earlier studies that most of the existing wire line and wireless homogeneous schedulers do not perform very well with respect to different traffic classes defined in WiMAX. In addition, each of this traffic classes has a different scheduling requirement and consequently, it has become necessary to design appropriate hybrid scheduling frameworks. Therefore, we propose an efficient hybrid packet scheduling scheme for IEEE 802.16 WiMAX to satisfy both delay and throughput guarantees for the admitted connections. An architecture model was developed to demonstrate the performance of the proposed scheme. Simulation results show that the proposed scheme is the best choice for QoS scheduling in WiMAX in terms of delay, throughput, bandwidth utilization and fairness of all connections of the system compared to schemes proposed in [11]. Simulation results also prove that the BS scheduler can guarantee the minimum bandwidth for each service flow and ensure fairness and QoS in distributing excess bandwidth among all connections. At the same time, the scheduler in SS can provide differentiated and flexible QoS support for all of the four scheduling service types. It can both reduce the delay of real-time applications and guarantee the throughput of non-real-time applications also enhancing the bandwidth utilization of the system and fairness index of the resources even at lower traffic intensity. Thereby eliminate starvation problem of lower priority class services. Therefore, the proposed QoS scheduling architecture can provide tight QoS guarantees for all types of traffic classes as defined in the scheduler standard.

## Acknowledgement

The authors deeply acknowledge the support from DST, Govt. of India for this work in the form of FIST 2007 Project on “Broadband Wireless Communications” in the Department of ETCE, Jadavpur University.

## BIOGRAPHIES

1. *Mr. Prasun Chowdhury* (prasun.jucal@gmail .com) has completed his Masters in Electronics and Telecommunication Engineering from Jadavpur University, Kolkata, India in 2009. Presently he is working as Senior Research Fellow (SRF) and also pursuing his PhD in the Department of Electronics and Telecommunication Engineering, Jadavpur University, Kolkata, India. His current research interests are in the areas of Call Admission Control and Packet Scheduling in IEEE 802.16 BWA Networks.

2. *Dr. Iti Saha Misra* (itimisra@cal.vsnl.net.in) is presently holding the post of Professor in the Department of Electronics and Telecommunication Engineering, Jadavpur University, Kolkata, India. After the completion of her PhD in Engineering in the field of Microstrip Antennas from Jadavpur University (1997), she is actively engaged in teaching since 1997. Her current research interests are in the areas of Mobility Management, Network Architecture and protocols, Integration Architecture of WLAN and 3G Networks, Call Admission control and packet scheduling in cellular and WiMAX networks, Location Management for Cellular Wireless Networks. Her other research activities are related to Design Optimization of Wire Antennas using Numerical Techniques like GA, PSO and BFA. She has authored more than 100 research papers in refereed Journal and International Conference and published a Book on Wireless Communications and Networks, by McGraw Hill, India. She is the recipient of the Career award for Young teachers by All India Council for Technical Education (AICTE) for the financial year 2003-2004 and obtained the IETE Gowri memorial award in 2007 for being the best paper in the general topic of “4G networks: Migration to the Future”. She has developed the OPNET, QualNet and VoIP laboratories in the Department of Electronics and Telecommunication Engineering of Jadavpur University to carry out advanced research work in Broadband wireless domain. She is the Senior Member of IEEE, founder Chair of the Women In Engineering, Affinity Group, IEEE Calcutta Section.